\documentclass[a4paper,11pt]{article}
\usepackage{pos}

\title{Status of the PrimEx $\eta$ experiment at Jefferson Lab}

\author*[a,1]{Alexander Somov}

\affiliation[a]{Thomas Jefferson National Accelerator Facility,\\
  Newport News, VA 23606, USA}

\emailAdd{somov@jlab.org}

\note{For the PrimEx $\eta$ working group and GlueX Collaboration}

\abstract{
The GlueX detector in the experimental Hall $D$ at Jefferson Lab offers a unique opportunity to perform a measurement of the decay width of eta mesons through the Primakoff effect. The PrimEx $\eta$ experiment complements the physics program at Jefferson Lab on measuring the decay width of light pseudoscalar mesons via the Primakoff process. The goal of PrimEx $\eta$ is to measure differential cross sections of $\eta$ mesons at forward angles using a beam of tagged photons incident on a liquid ${}^{4}{\rm He}$ target. The data will be used for the extraction of the decay width. This measurement is vital for understanding fundamental properties like the ratios of the light quark masses and the $\eta$-$\eta^\prime$ mixing angle, and will provide an important test of chiral symmetry breaking in QCD. Our experimental results will help reduce uncertainties on partial widths of all other $\eta$ decays. The experiment collected data during three physics runs between 2019 and 2022. We will give an overview of the PrimEx $\eta$ experiment and the current status of our data analyses. We will also discuss the feasibility of conducting future Primakoff measurements in light of the recent upgrade of the GlueX forward calorimeter and the potential accelerator energy upgrade to 22 GeV. }

\FullConference{
The 11th International Workshop on Chiral Dynamics (CD2024)\\
 26-30 August 2024\\
Ruhr University Bochum, Germany\\ }

\begin{document}
\maketitle

\section{Introduction}
Secondary photon beams provided at Jefferson Lab (JLab) offer a unique capability to study the production of neutral pseudoscalar mesons in the Coulomb field of nuclei. This process is known as the Primakoff effect. The Primakoff program at JLab began in 2004 with the first Primakoff experiment~\cite{PrimEx-II:2020jwd,PrimEx:2010fvg} in experimental Hall B, using beam photons with a maximum energy of 6 GeV. This experiment performed a precision measurement of the differential cross section of $\pi^0$ mesons and extracted the radiative decay width $\Gamma(\pi^0\to\gamma\gamma)$. Following an upgrade of the Continuous Electron Beam Accelerator Facility (CEBAF), which increased the beam energy to 12 GeV, and the construction of the new experimental Hall D with the GlueX detector, a new experiment, PrimEx $\eta$, was proposed. This experiment aimed to measure the decay width of $\eta$ mesons, specifically $\eta \to \gamma \gamma$. The experiment successfully collected data in 2019, 2021, and 2022.

The GlueX detector has also been used to search for the production of axion-like particles via the Primakoff process~\cite{Pybus:2023yex}. These particles predominantly couple to photons. In addition, the GlueX detector has been employed in experiments to measure the charged and neutral pion polarizabilities in the reactions $\gamma \gamma \to \pi^+ \pi^-$ and $\gamma \gamma \to \pi^0 \pi^0$, respectively~\cite{ch_pol, neut_pol}.

In this paper, we provide an overview of the PrimEx $\eta$ experiment. The paper is organized as follows: Sections 2 and 3 describe the physics motivation and the Primakoff method, Sections 4 and 5 cover the experiment and the status of the analyses, and Section 6 discusses the upgrade of the GlueX forward calorimeter, future plans for $\eta$ measurements, and the feasibility of extending the Primakoff program with the GlueX detector for a potential future accelerator energy upgrade to 22 GeV.

\section{Physics motivation}
Studying the properties of $\pi^0$, $\eta$, and $\eta^\prime$ mesons is important for understanding the fundamental symmetry structure of low-energy QCD. The spontaneously broken $SU_{\rm L}(3)\times SU_{\rm R}(3)$ symmetry to the flavor $SU(3)$, caused by the condensation of quark-antiquark pairs in the QCD vacuum, generates eight massless pseudoscalar Goldstone bosons. These can be represented as an octet of mesons: $\pi^0$, $\pi^\pm$, $K^{\pm}$, $K^0$, and $\eta$. The non-zero quark masses explicitly break the chiral symmetry and give rise to the masses of the Goldstone bosons. The heaviest octet member, $\eta$, is of particular interest because its specific and strong decays are suppressed by the conservation of P, PC, C, and G-parity symmetries, as well as angular momentum, resulting in a significantly smaller decay width. This makes $\eta$ more sensitive for testing higher-order Chiral Perturbation Theory (ChPT) predictions and for searches for physics beyond the Standard Model.

Studying the decay widths of the $\eta$ and $\eta^\prime$ mesons is of particular importance as it provides critical insights into the mixing between the $\eta$ and $\eta^\prime$ mesons and contributes to the extraction of low-energy constants in ChPT. The two-photon decay of $\eta$ mesons originates from the chiral anomaly~\cite{Bell:1969ts, Adler:1969gk}, which is related to the breaking of the classical symmetry by quantum fluctuations of the quark fields that couple to a gauge field (photons). The $\eta$ decay width can be theoretically predicted by ChPT in the chiral limit and the large number of colors, $N_c$, limit. The flavor $SU(3)$  and isospin breaking due to unequal quark masses results in mixing among the $\pi^0$, $\eta$, and $\eta^\prime$ mesons. The physical $\eta$ and $\eta^\prime$ states are represented as compositions of the pure $SU(3)$ states, the octet $\eta_8$ and the singlet $\eta_0$, related to each other through mixing angles, which are the fundamental low-energy constants in ChPT. The two-photon decay widths of the $\eta$ and $\eta^\prime$ mesons can be expressed in terms of these fundamental parameters and extracted from measurements. Therefore, the decay widths of the $\eta$ and $\eta^\prime$ mesons must be analyzed together.
\begin{figure}[t]
\begin{center}
\includegraphics[width=0.55\linewidth,angle=0]{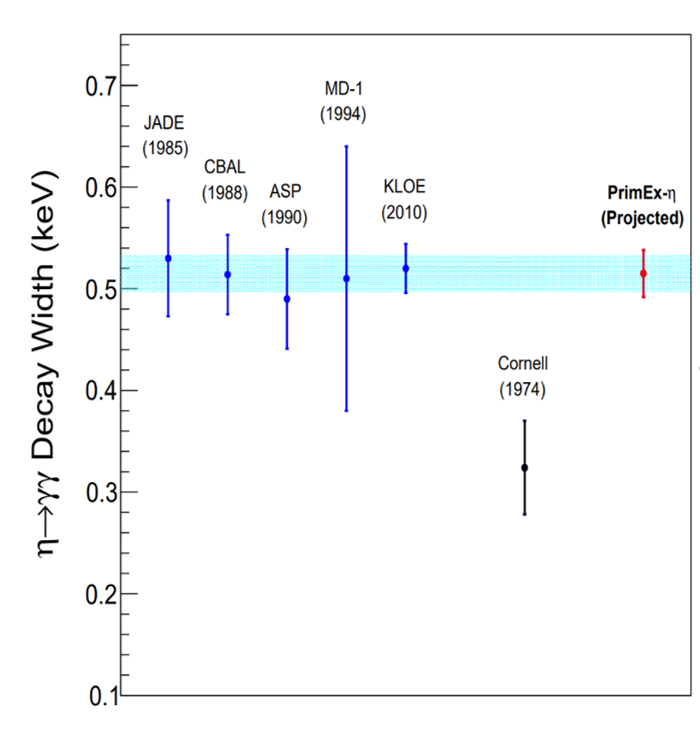}
\end{center}
\caption{Measurements of the radiative decay width of $\eta$ meson and the projected result from the PrimEx $\eta$ experiment. Note that the actual uncertainties in the PrimEx measurement may be 2 to 2.5 times larger due to the beamline background observed during the experiment.}
\label{fig:width_measurements}
\end{figure}
The precise determination of the $\eta$ decay width into two photons, would help to improve the determination of the other partial decay widths of the $\eta$ meson. Most partial decay widths are not measured directly, but calculated using the two-photon width and the respective branching ratios. These would be particularly beneficial for $\eta\to 3\pi^0$ and $\eta\to\pi^+\pi^-\pi^0$ decays. Those are strong decays, which proceed from the isospin breaking due to the unequal masses of the $u$ and $d$ quarks. The amplitudes of these decays depend on the quark mass ratio 
\begin{equation}
Q^{2}_{\rm m}=\frac{m^2_s-\hat{m}^2}{m^2_d-m^2_u},
\end{equation}
where $\hat{m} = (m_u+m_d)/2 $. The extraction of the quark mass ratio from isospin breaking is described in Ref.~\cite{Bijnens:2002qy,Anisovich:1996tx}.

The radiative decay width of $\eta$ meson has been measured in two classes of experiments: using the Primakoff method in the reaction $\gamma + A \to \eta + A$ and in the collider experiments using the process $e^+ e^-\to e^+e^-\gamma^\star\gamma^\star \to e^+e^-\eta$. The first measurement was performed by a group at Cornell University in 1974, using the Primakoff method, which will be described in Section~\ref{sec:primakoff_method}. The group measured the differential photoproduction cross section on $\eta$ mesons using 5 nuclear targets and three beam energies: 5.8 GeV, 9.0 GeV, and 11.45 GeV~\cite{Browman:1974sj}. The fixed target measurement was followed by 5 collider experiments at PETRA, DESY, SLAC, VEPP-4, and DA$\Phi$NE electron-positron colliders~\cite{ParticleDataGroup:2024cfk}. The previous measurements of the decay width along with a projection of the PrimEx $\eta$ experiment are presented in Fig.~\ref{fig:width_measurements}. A comparison of the measurements demonstrates a discrepancy between the Primakoff and collider results. The PrimEx $\eta$ experiment is expected to shed light on this long-standing discrepancy and improve the overall experimental uncertainty.
\begin{figure}[t]
\begin{center}
\includegraphics[width=0.4\linewidth,angle=0]{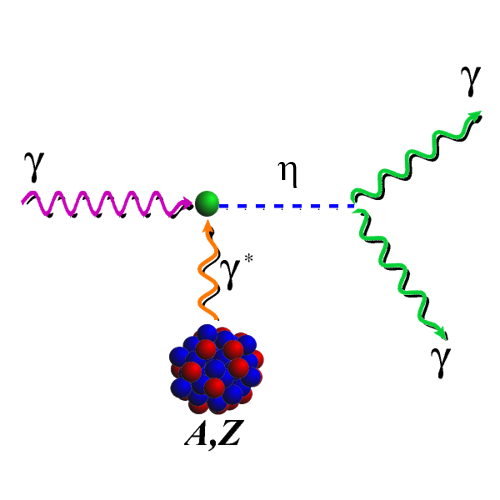}
\end{center}
\caption{Photoproduction of an $\eta$ meson by a photon in the Coulomb field of a nucleus (Prmakoff effect).}
\label{fig:primakoff_diagram}
\end{figure}
\section{The Primakoff method} 
\label{sec:primakoff_method}
The production of pseudoscalar mesons in the Coulomb field of a nucleus by real photons, presented in Fig.~\ref{fig:primakoff_diagram}, is referred to as the Primakoff process. The Primakoff production cross section of $\eta$ mesons, which decay into two photons, by unpolarized photons on a zero-spin nuclear target is given by~\cite{Bellettini:1970th}:
\begin{equation}
\frac{d\sigma_{\rm P}}{d\Omega}=\Gamma_{\gamma\gamma}\frac{8\alpha Z^2}{m^3_\eta}\frac{\beta^3 E^4}{Q^4}|F_{e.m.}(Q)|^2{\rm sin}^2\theta,
\label{eq:prim}
\end{equation}
where $\Gamma_{\gamma\gamma}$ and $m_\eta$ are the decay width and mass of the $\eta$, $E$ is the energy of the incoming photon, $\alpha$ is the QED coupling, $\beta$ and $\theta$ are the velocity and production angle of the meson, $Q$ is the four-momentum transferred to the nucleus, and $Z$ and $F_{\rm e.m}$ are the atomic number and the electromagnetic form factor of the nucleus, respectively.

\begin{figure}[t]
\begin{center}
\includegraphics[width=0.6\linewidth,angle=0]{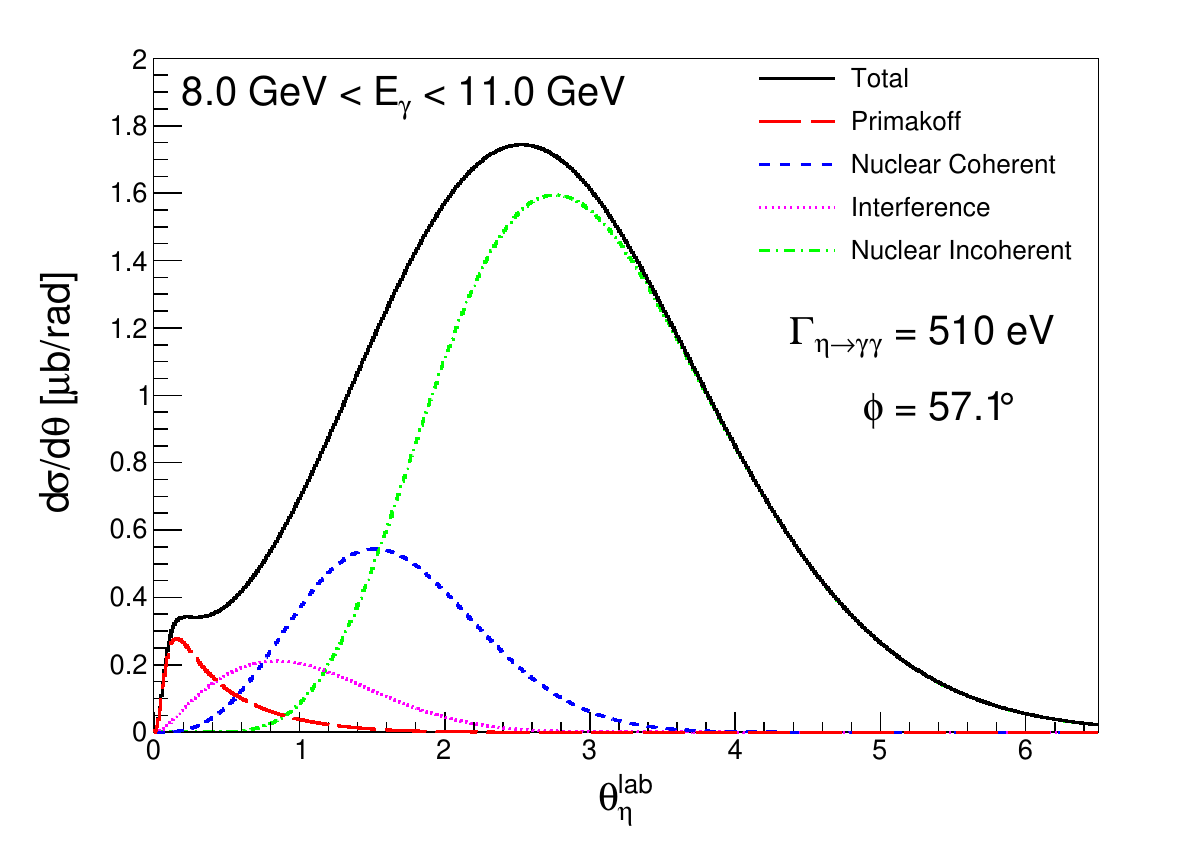}
\end{center}
\caption{Theoretical calculations of the differential cross section for $\eta$ meson Primakoff production and associated backgrounds on a ${}^4$He target~\cite{primex}.
The long dashed line  corresponds to the Primakoff process, the dashed line is the nuclear coherent  process, the dash-dotted line is the incoherent process, the dotted line is the interference between the Primakoff and nuclear coherent, and the solid line is the total.}
\label{fig:primakoff_mc}
\end{figure}

The Primakoff production must be separated from backgrounds originating from $\eta$ produced in the  interaction between photons and the nuclear hadronic field, including both coherent and incoherent productions of $\eta$ mesons, as well as interference between the nuclear hadronic  and Primakoff production amplitudes. The total photoproduction cross section for $\eta$ mesons can be written as
\begin{equation}
\frac{d\sigma}{d\Omega}=\frac{d\sigma_{\rm P}}{d\Omega}+\frac{d\sigma_{\rm C}}{d\Omega}+\frac{d\sigma_{\rm I}}{d\Omega} + 2\sqrt{\frac{d\sigma_{\rm P}}{d\Omega}\frac{d\sigma_{\rm C}}{d\Omega}} {\rm cos}\phi,
\label{eq:prim_total}
\end{equation}
where $\frac{d\sigma_{\rm C}}{d\Omega}$ and $\frac{d\sigma_{\rm I}}{d\Omega}$ are the coherent and incoherent cross sections and $\phi$ is the interference angle. The nuclear coherent cross section can be expressed as:
\begin{equation}
\frac{d\sigma_{\rm C}}{d\Omega}=C\cdot {\rm sin}^2\theta \cdot A^2 \cdot |F_N(Q)|^2,
\end{equation}
where $|F_N(Q)|$ is the nuclear form factor corrected for the nuclear absorption of $\eta$ mesons, $A$ is the nucleon number, and $C\cdot {\rm sin}^2 \theta$ arises from the isospin and spin-independent part of the photoproduction amplitude on a single nucleon. The incoherent cross section can in general be written as:
\begin{equation}
\frac{d\sigma_{\rm I}}{d\Omega}=\frac{d\sigma_{\rm N}}{d\Omega}\cdot A\cdot \xi\cdot (1-G(Q)),
\end{equation}
where $\frac{d\sigma_{\rm N}}{d\Omega}$ represents the $\eta$ photoproduction cross section on a single nucleon, $\xi$ is the absorption factor, and  $(1-G(Q))$ is a factor, that ensures the cross section approaches zero at small momentum transfer due to the Pauli exclusion principle. The nature of incoherent production is more complex and depends on the specific theoretical models. Fig.~\ref{fig:primakoff_mc} shows theoretical calculations of the differential cross section for $\eta$ meson Primakoff production and associated backgrounds on a $^4$He target, with beam photon energies ranging from 8 GeV to 11 GeV.

Several criteria must be considered when planning Primakoff experiments. As indicated by Eq.~(\ref{eq:prim}), the Primakoff cross section increases with the beam energy. Experiments conducted at higher energies have the advantage of better separation between the Primakoff peak and background. This will be further discussed in Section~\ref{sec:22_gev}. The cross section strongly depends on the atomic number (as $Z^2$), which favors the use of heavy nuclear targets. However, for large values of $A$, the coherent background peaks close to the Primakoff peak, complicating the separation of signal from background, especially from the interference term. In contrast to the Primakoff production of $\pi^0$ mesons, $\eta$ mesons have a much larger mass. As a result, their production requires a significantly larger momentum transfer for a given production angle. Uncertainties from nuclear excitations may complicate the enhancement of coherence in Primakoff production when using heavier targets, compared to more bound light targets. Recent theoretical calculations specifically performed for the PrimEx group look promising~\cite{Fix:2023vzj} and will be considered for the potential extension of the PrimEx physics program using heavier targets. This will be further discussed in Section~\ref{sec:alex_fix}. In the PrimEx $\eta$ experiment, a liquid ${}^4$He target was used. The beam energy range used in data analyses was between 8 GeV to 11.7 GeV.
\begin{figure}[t]
\begin{center}
\includegraphics[width=0.7\linewidth,angle=0]{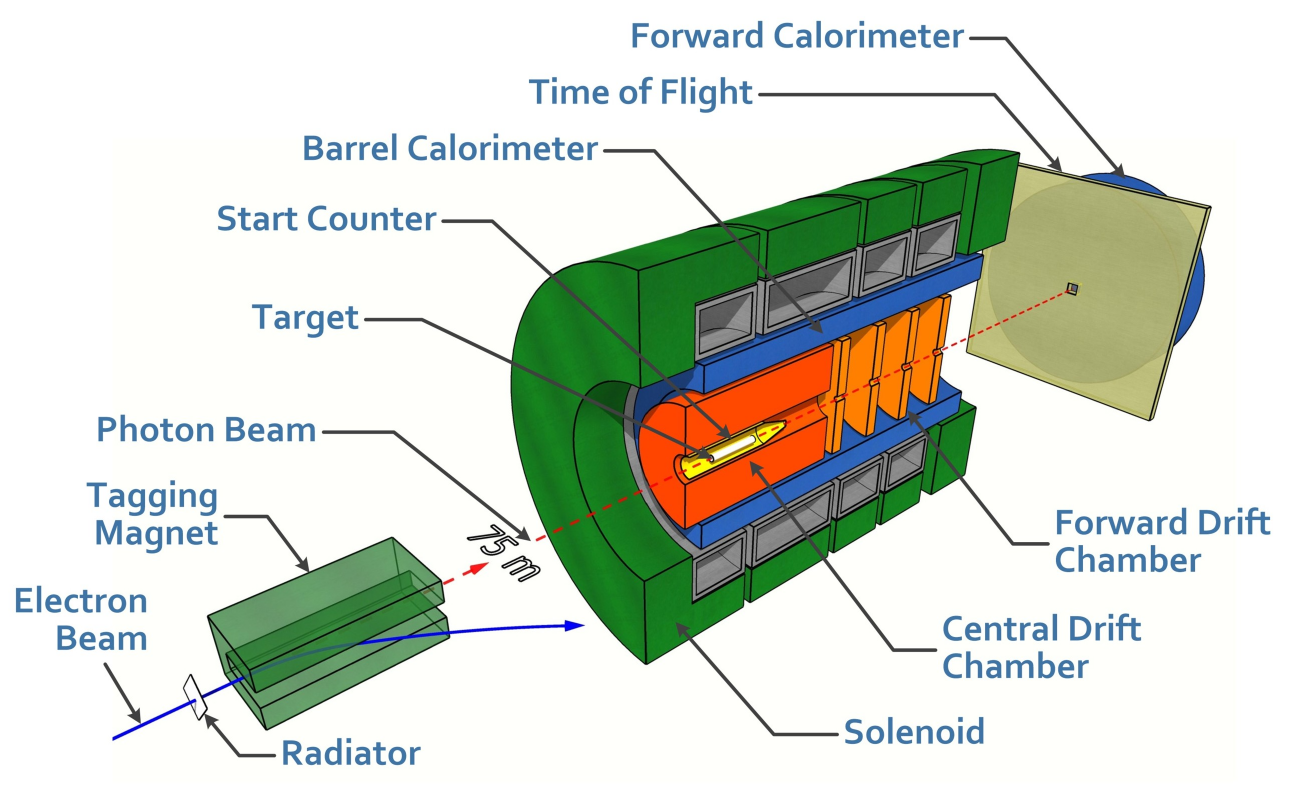}
\end{center}
\caption{Schematic view of the GlueX forward spectrometer and the tagger system.}
\label{fig:gluex_detector}
\end{figure}
\begin{table}[ht]
\begin{center}
\renewcommand{\arraystretch}{1.4}
\begin{tabular}{lccc}
\hline 
       &  Phase I  &  Phase II  & Phase III \\
\hline       
Year   &  2019     &   2021     &   2022   \\
$E_{\rm beam}$ (GeV)  &  11.2  &  10.0  &  11.6  \\
Luminosity $({\rm pb}^{-1})$ &  6   &  2  &  17  \\
Magnetic field   & Off  &  Off (most runs)  & On \\
\hline
\end{tabular}
\end{center}
\caption{Run conditions of the PrimEx $\eta$ experiment and the integrated luminosity in the beam energy range $E_\gamma \ge 8\;{\rm GeV}$. $E_{\rm beam}$ is the energy of the electron beam used to produce the photon beam.}
\label{tab:primex_exp}
\end{table}

\section{PrimEx $\eta$ experiment in Hall D}

The physics goal of the PrimEx $\eta$ experiment~\cite{primex} is to measure the $\eta \to \gamma \gamma$ decay width via the Primakoff effect. This decay width is extracted from the measurement of the photoproduction cross section of $\eta$ mesons in the Coulomb field of a nucleus, as described in Section~\ref{sec:primakoff_method}. The measurements were conducted using the GlueX detector, located in experimental Hall D at Jefferson Lab, which was specifically designed for experiments utilizing a photon beam.

Beam photons are produced by an electron beam from JLab's continuous electron beam accelerator facility, which can deliver a maximum energy of 12 GeV. Electrons from the accelerator pass through a thin radiator inserted into the beam, generating photons via the bremsstrahlung process. The energy of a beam photon, $E_{\gamma}$, is determined by measuring the energy of the scattered bremsstrahlung photon, $E_{\rm scat}$, in the Hall D tagging system, and knowing the energy of the initial electron, $E_{\rm beam}$, before radiating the photon. This relationship is given by $E_{\gamma}$ = $E_{\rm beam}$ - $E_{\rm scat}$. The typical energy resolution of the beam photon is about $0.1\%$. The beam photons are sent to the target of the GlueX detector presented in Fig.~\ref{fig:gluex_detector}.

The GlueX detector~\cite{gluex} is a forward spectrometer consisting of a solenoid magnet located around the target, capable of generating a magnetic field of about 2 T. Charged tracks are reconstructed in the central and forward drift chambers. Reconstruction of photons is performed in the forward (FCAL) and barrel (BCAL) electromagnetic calorimeters. The particle identification is performed in the Time-Of-Flight (TOF) scintillator detector, which is positioned in front of the forward calorimeter and measures the time of flight of charged particles. In the Primakoff analysis the detector is used to veto charged tracks. 

$\eta$ mesons are produced inside a 29.5-cm-long ($\sim 3.9\%$ radiation length) cryogenic liquid $^4$He target. The density of the target during the run was monitored by reconstructing Compton scattering events. Special calibration runs were conducted at the beginning of the experiment using a beryllium foil target with a well-known density of about $5\%$ radiation length. To measure the background originating in the photon beamline between the target and the forward calorimeter, we also collected data when the target cell was not filled with liquid helium.

Photons originating from the $\eta\to\gamma\gamma$ decays were reconstructed in the forward calorimeter, which is situated about 6 m downstream the beam from the target.  The calorimeter consists of 2800 rectangular lead glass blocks with a size of 4 cm $\times$ 4 cm $\times$ 45 cm. It has a 12 cm $\times$ 12 cm hole in the middle for the beam, and covers a polar angle between $1^\circ$ and $11^\circ$ for photons originating from the target. The typical energy resolution of the lead glass calorimeter is $\sigma_{\rm E}/E = 6.2\%/\sqrt E \oplus 4.7\%$~\cite{gluex}.

\begin{figure}[t]
\begin{center}
\includegraphics[width=0.65\linewidth,angle=0]{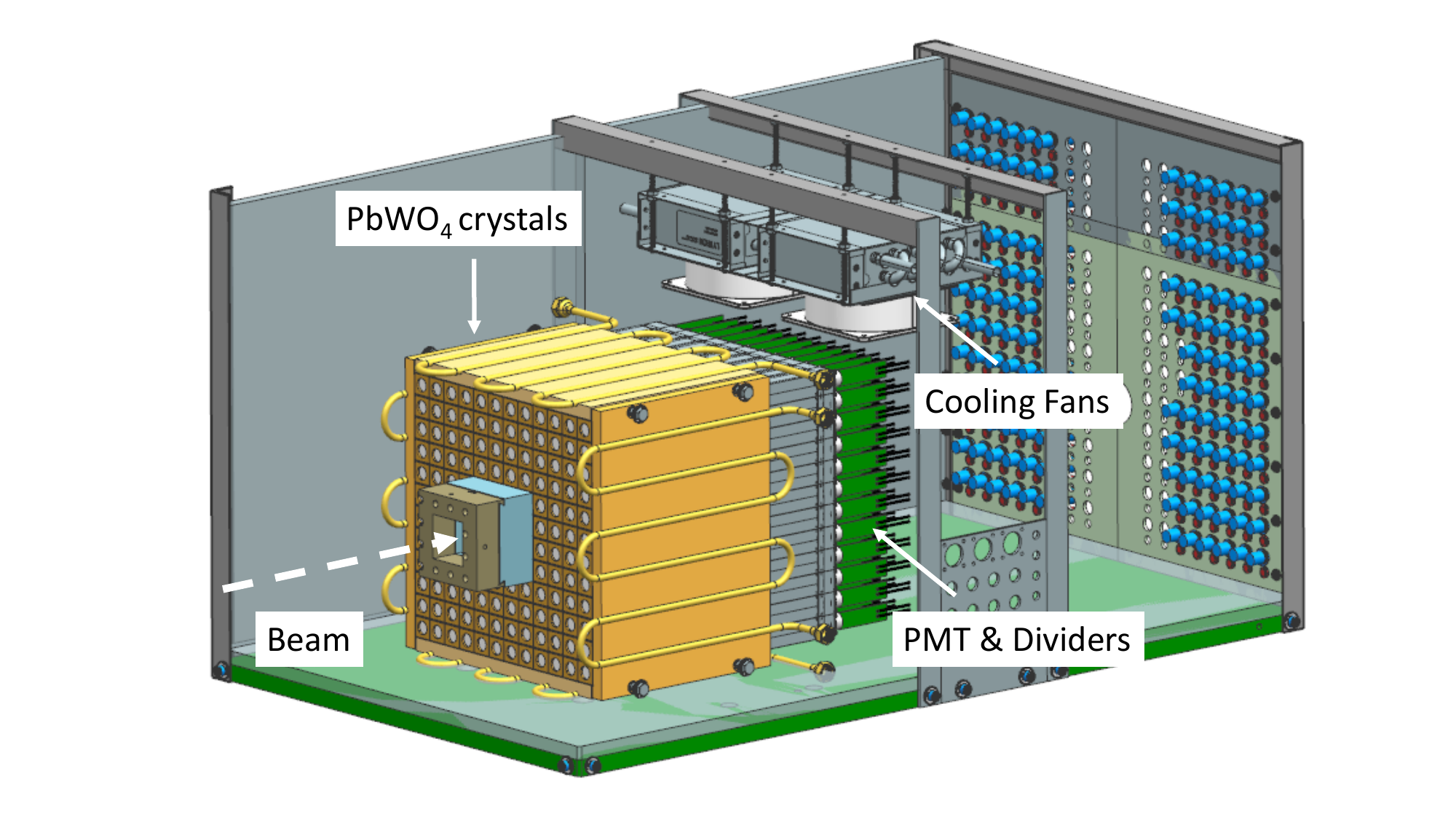}
\end{center}
\caption{Schematic view of the PrimEx Compton calorimeter consisting of an array of $12 \times 12$ PbWO$_4$ scintillating crystals with a beam hole of $2 \times 2$ crystals in the middle~\cite{ccal}.}
\label{fig:ccal}
\end{figure}

To ensure detector stability and monitor luminosity, we constructed a compact calorimeter, referred to as the Compton Calorimeter (CCAL)~\cite{ccal}. The CCAL comprises 140 lead tungstate scintillating crystals, each measuring  2 cm $\times$ 2 cm $\times$ 20 cm. This calorimeter was positioned approximately 6 m downstream from the FCAL. It enabled the reconstruction of electrons and photons originating from Compton scattering events produced in the target at small angles, where particles could not be detected by the FCAL as they passed through the hole around the beam pipe. The calorimeter extended the polar angle coverage of the detector to small angles, down to about $0.2^\circ$. It provides a factor of four better granularity and about a factor of two better energy resolution compared to the FCAL. A schematic view of the CCAL is shown in Fig.~\ref{fig:ccal}. Both calorimeters were included into the GlueX trigger system, which was based on the energy depositions in the FCAL and CCAL. 

The PrimEx run was organized into three phases. The first data set was collected in 2019 with the solenoid magnetic field turned off. During this phase, the CCAL was commissioned, and data were collected for precise measurements of the Compton scattering cross section, as well as for analyses of the $\eta \to \gamma \gamma$ and $\eta \to 3\pi^0$ decays. Note that the magnetic field deflects electrons in the Compton scattering process, complicating the analysis. The impact of the magnetic field was studied during the PrimEx Phase II run in 2021 when the field was turned on for some period of the run. In Phase III (2022), the magnetic field was turned on, which enabled the inclusion of the charged $\eta$ decay, $\eta \to \pi^+ \pi^- \pi^0$, into the Primakoff program. The accelerator beam energy varied during these three phases, with a maximum energy of 11.6 GeV achieved in 2022. The run conditions of the PrimEx $\eta$ experiment and the corresponding integrated luminosity are listed in Table~\ref{tab:primex_exp}.

\section{Status of data analyses}
The PrimEx $\eta$ experiment completed its data-taking in 2022 and is currently calibrating and processing the data acquired during experimental Phase III. Below, we describe the status of the measurement of the Compton scattering cross section and provide an overview of the Primakoff analysis of $\eta \to \gamma \gamma$ decays.
\subsection{Compton scattering cross section}
\begin{figure}[t]
\begin{center}
\includegraphics[width=0.65\linewidth,angle=0]{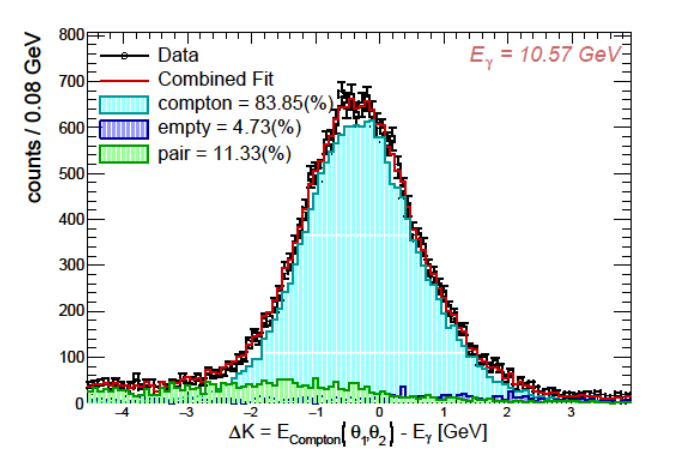}
\end{center}
\caption{$\Delta K$ distribution from the Compton analysis.}
\label{fig:dk}
\end{figure}

\begin{figure}[t]
\begin{center}
\hspace{0.2in}
\includegraphics[width=0.67\linewidth,angle=0]{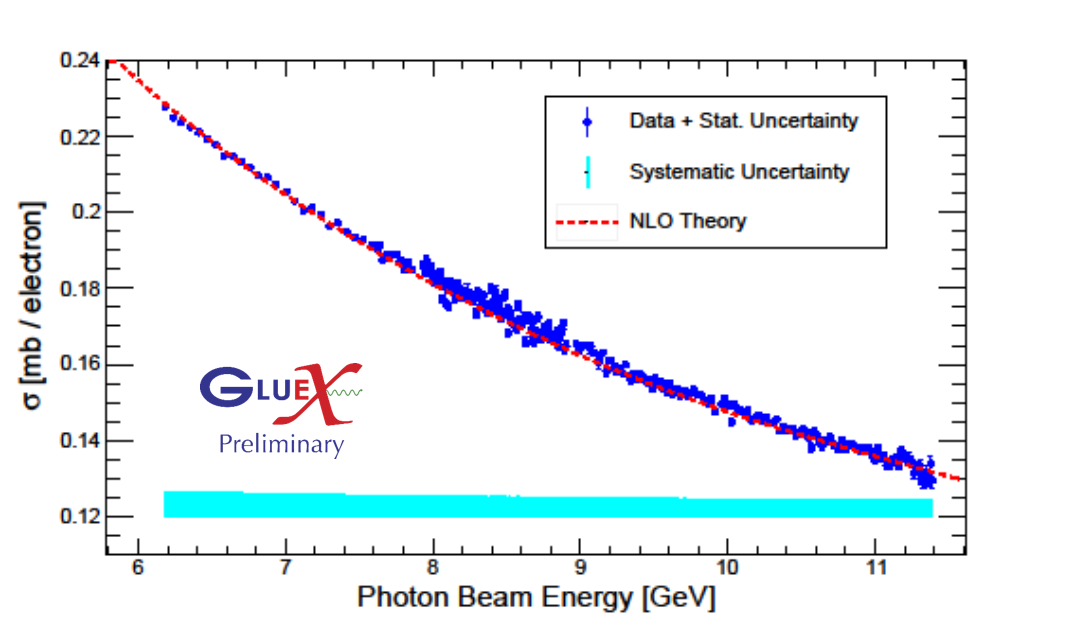}
\end{center}
\caption{Preliminary Compton cross section measurements on the Be target as a function of the beam energy.}
\label{fig:comp_cross_sect}
\end{figure}

Instrumenting the GlueX detector with the Compton calorimeter provides a unique capability to perform the first measurement of the Compton scattering cross section in the energy range between 6.5 GeV and 11 GeV. This energy range is covered by the pair spectrometer~\cite{Barbosa:2015bga}, which is used in the experiment to measure the flux of beam photons. The data sample was collected using a beryllium target, with the Solenoid magnetic field turned off. Compton candidates were reconstructed using the well-defined kinematics of the two-body reaction. Events with two reconstructed showers, one in the FCAL and the other in the CCAL, originating from the same beam interaction, were selected. The event selection criteria included requirements on the coplanarity of the reconstructed photon and electron, as well as elasticity, defined as the difference between the reconstructed energy in the event and the beam energy. The energy resolution of reconstructed Compton candidates in this energy range is 120 MeV. The primary background originates from the pair production reaction. The Compton yield was obtained by fitting the $\Delta K$ distribution defined as $\Delta K = E_{\rm COM}(\theta_{\rm CCAL},\theta_{\rm FCAL})-E_\gamma$, where $E_{\rm COM}(\theta_{\rm ECAL},\theta_{\rm CCAL})$ is the initial beam photon energy, computed using the shower's polar angle and the Compton kinematics. This variable helps to better distinguish the pair production background from the signal. An example of the $\Delta K$ distribution is presented in Fig.~\ref{fig:dk}. The preliminary Compton cross section as a function of the beam energy, measured on the Be target, is shown in Fig.~\ref{fig:comp_cross_sect}. More details regarding the analysis can be found in Ref.~\cite{Smith:2024wvr}. The analysis results are currently being finalized and prepared for a journal publication.

\begin{figure*}[]
\centering
\vspace{0.1in}
$\vcenter{\hbox{\vspace{-0.12in}\includegraphics[width=0.54\linewidth]{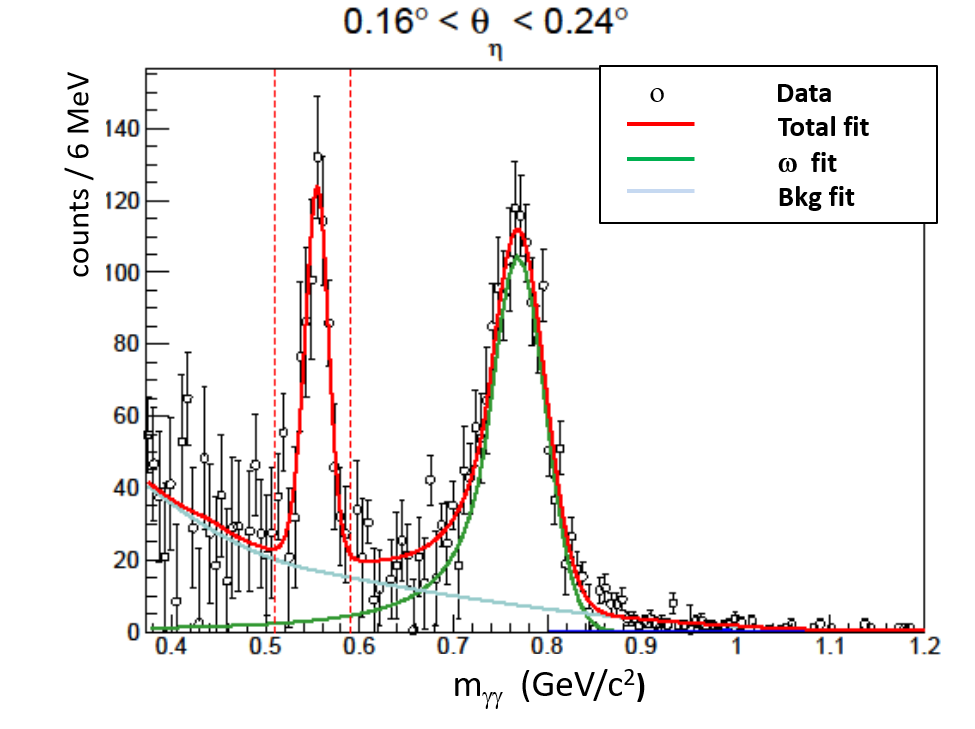}}}$
$\vcenter{\hbox{\includegraphics[width=0.43\linewidth]{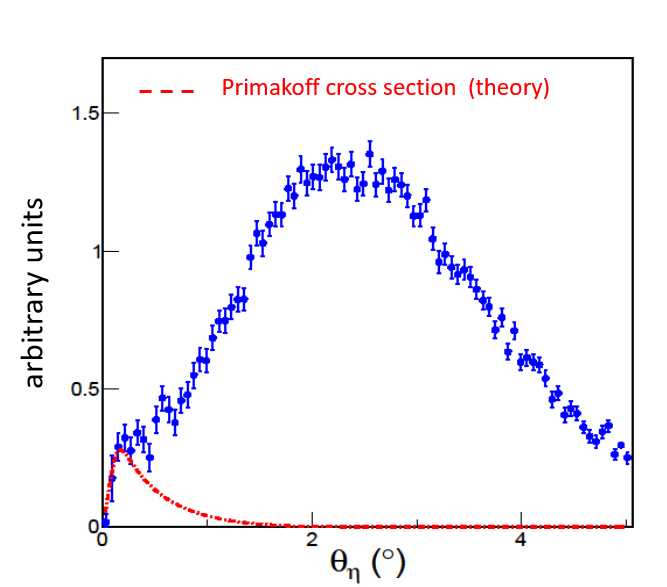}}}$
\caption{The invariant mass distribution of two reconstructed photons, with the  solid lines representing the fit results. The shape of the differential cross section obtained from the analysis of the PrimEx Phase I data (right). Theoretical predictions of the Primakoff cross section are superimposed on this plot. }
\label{fig:lineshape}
\end{figure*}

\subsection{Primakoff analysis}
To measure the differential cross section for the Primakoff production of $\eta$ mesons in the reaction $\gamma + {}^4\text{He} \to \eta + {}^4\text{He}$ ($\eta \to \gamma \gamma$), we selected two photons in the FCAL, originating from the same beam bunch. To reduce contributions from the inclusive production of $\eta$ mesons and background from other hadronic interactions, we rejected events with extraneous photons in the BCAL and FCAL, as well as tracks identified using hits in the Time-of-Flight detector.

In addition to the main physics backgrounds described in Section~\ref{sec:primakoff_method}, we encountered a relatively large background (about $30\%$ of the signal at small angles) from beam interactions with air and beamline material downstream of the target. This background was measured using special runs with an empty target and then subtracted. The polar angle distribution of the $\eta$ candidates was divided into several bins. For each bin, we obtained the yield of $\eta$ mesons by fitting the di-photon invariant mass distribution, $m_{\gamma\gamma}$. An example of the fit to the invariant mass distribution is shown in the left plot of Fig.~\ref{fig:lineshape}. The $\eta$ meson lineshape was parameterized as a sum of Gaussian functions, while the background from the $\omega \to \pi^0 \gamma$ decay was modeled using a Crystal Ball function.\footnote{The function is named after the Crystal Ball collaboration.} An exponential function was applied to model the remaining background. The yield of $\eta$ candidates was then used to calculate the cross section. The preliminary shape of the differential cross-section, obtained from the PrimEx Phase I data, is shown in the right plot of Fig~\ref{fig:lineshape}. Theoretical predictions for the Primakoff cross section assuming the decay width of $\Gamma(\eta\to\gamma\gamma) = 510\;{\rm eV}$, are superimposed on this plot. Further details on the ongoing analysis can be found in Ref.~\cite{Smith:2024wvr}.

\section{Extension of the Primakoff program with the GlueX detector}
We are studying the feasibility of extending the Primakoff program with the GlueX detector beyond the ongoing PrimEx $\eta$ experiment. There are several ways of improving the measurements and extending the program for the future experiments. Below we list some of them:
\begin{itemize}
\item upgrade of the GlueX forward calorimeter
\item use heavier targets, measure the decay width of $\eta^\prime$ mesons
\item extend the Primakoff program for the potential upgrade of the JLab's energy to 22 GeV
\end{itemize}
\subsection{Upgrading forward calorimeter of the GlueX detector}
\begin{figure*}[]
\centering
$\vcenter{\hbox{\includegraphics[width=0.45\linewidth]{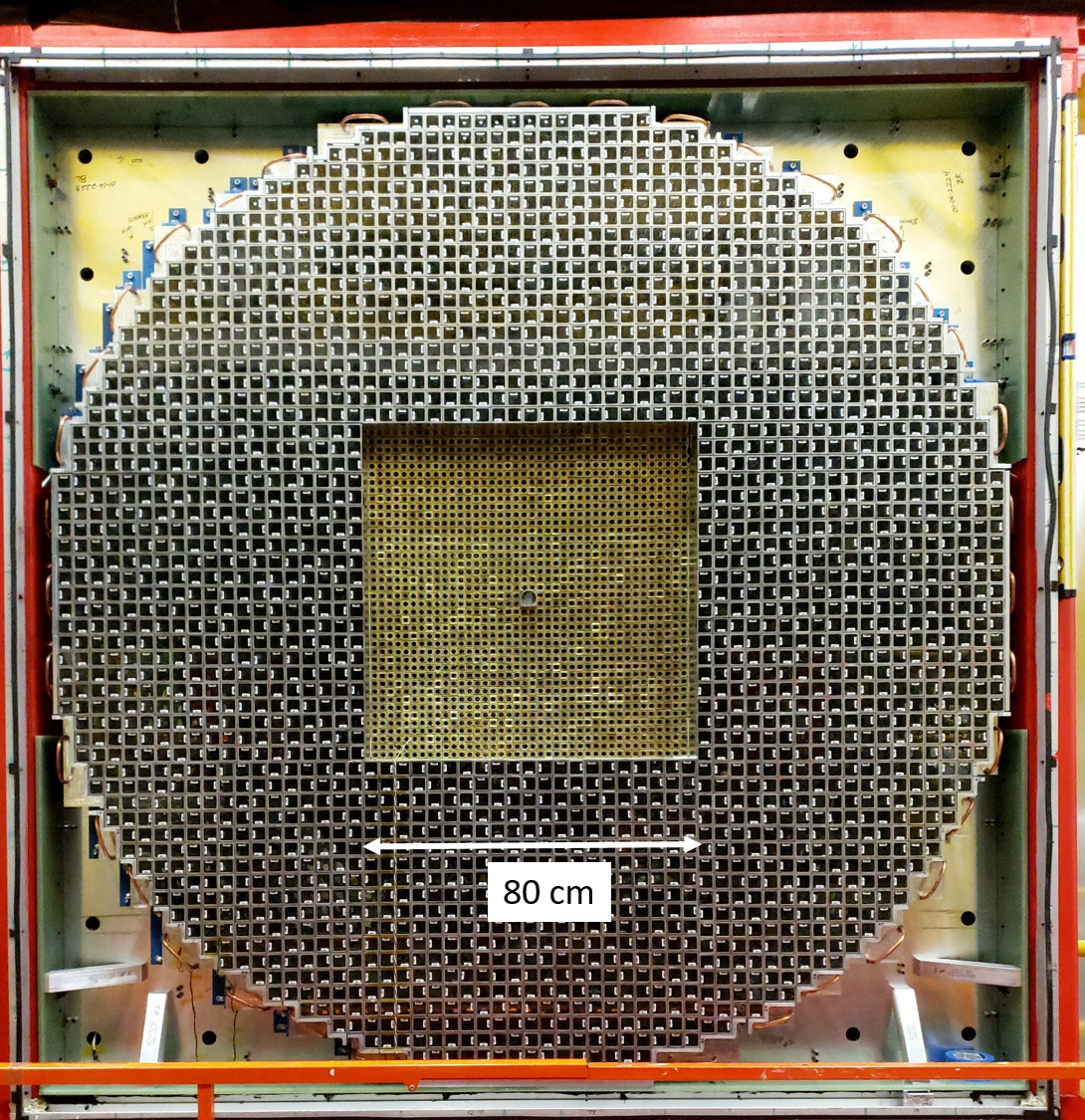}}}$
\hspace{0.2in}
$\vcenter{\hbox{\includegraphics[width=0.50\linewidth]{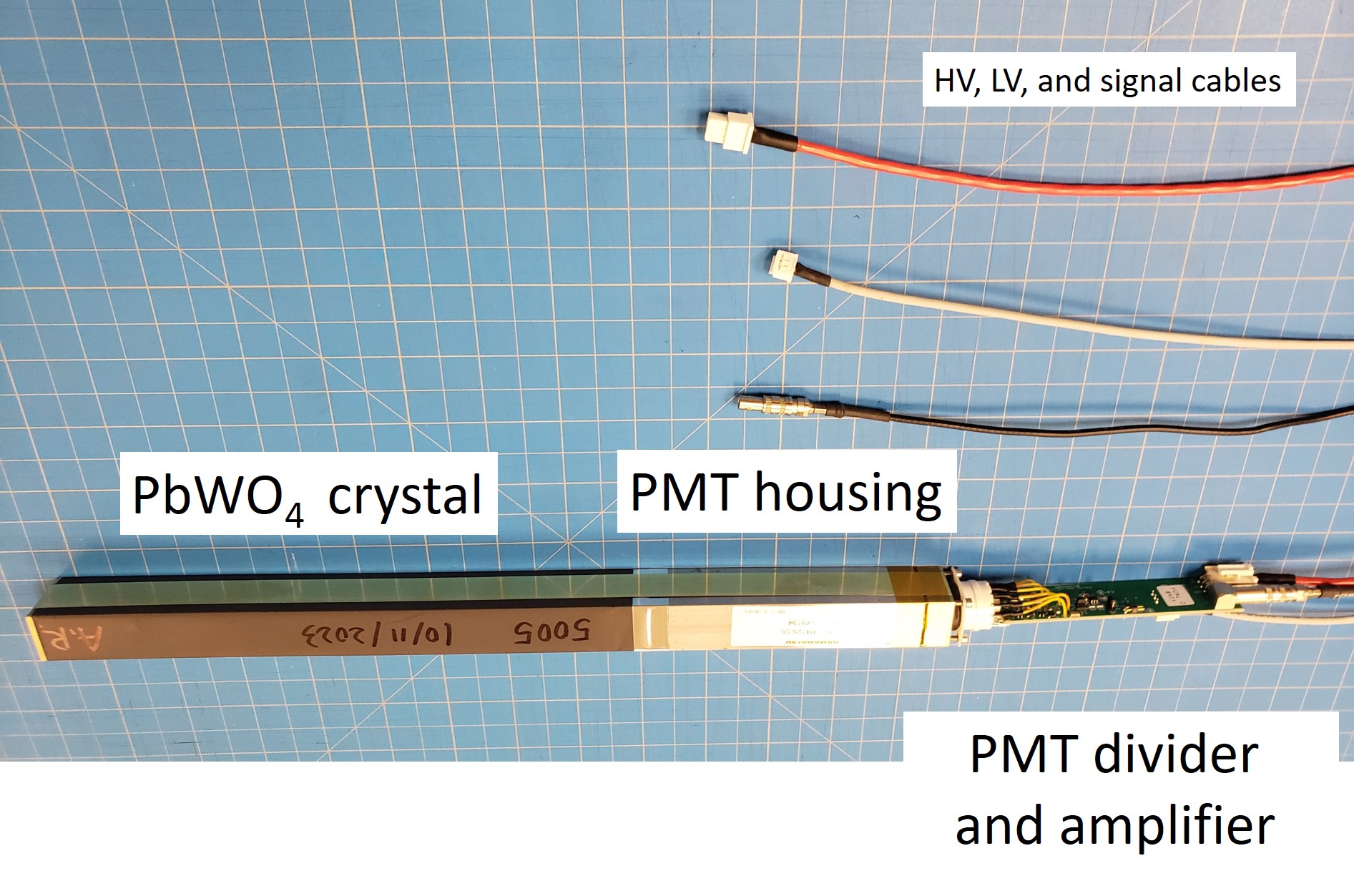}}}$
\caption{The upgraded forward calorimeter of the GlueX detector. The new lead tungstate calorimeter, ECAL, located at  the center, is surrounded by lead glass modules (left). An ECAL module with its main components (right).}
\label{fig:ecal}
\end{figure*}

The JLab Eta Factory (JEF) experiment~\cite{jef} required an upgrade to the inner part of the FCAL, replacing it with high-resolution, high-granularity scintillating crystals. The new electromagnetic calorimeter (ECAL) consists of 1596 scintillating crystals organized into a $40 \times 40$ array of modules. Each ECAL crystal is the same size as the crystals used in the CCAL. The new forward calorimeter of the GlueX detector was fabricated at JLab and installed in Hall D in 2025~\cite{ecal}. It is shown in Fig.~\ref{fig:ecal}. The energy resolution of the ECAL is approximately a factor of two better than that of the original calorimeter. This upgrade significantly improves the reconstruction of photons in the forward direction, as well as the energy and angular resolutions of reconstructed mesons, which is critical for the Primakoff program.
\subsection{Primakoff program using heavier nuclear target}
\label{sec:alex_fix}
Understanding the coherent and incoherent backgrounds and their separation from the signal is critical for the measurement of the Primakoff cross section. Since the Primakoff cross section depends on the $Z^2$ of the nucleus, it is advantageous to use a heavy target. However, the production of heavy mesons like $\eta$ and $\eta^\prime$ requires a large momentum transfer to the nucleus, leading to theoretical uncertainties due to the excitation of the nucleus to discrete bound-state levels. Recently, theoretical calculations were performed to evaluate the contributions of various 1$p$-shell level excitations to the differential cross section of incoherent production~\cite{Fix:2023vzj}. Contributions from various nuclear excitations to the cross section for ${}^{12}C(\gamma,\eta)^{12}C^\star$ are presented in Fig.~\ref{fig:alex_fix}. These contributions are relatively small compared with the Primakoff cross section and should be verified with experimental data.

\subsection{Primakoff measurements at 22 GeV}
\label{sec:22_gev}
\begin{figure}[t]
\begin{center}
\includegraphics[width=0.48\linewidth,angle=0]{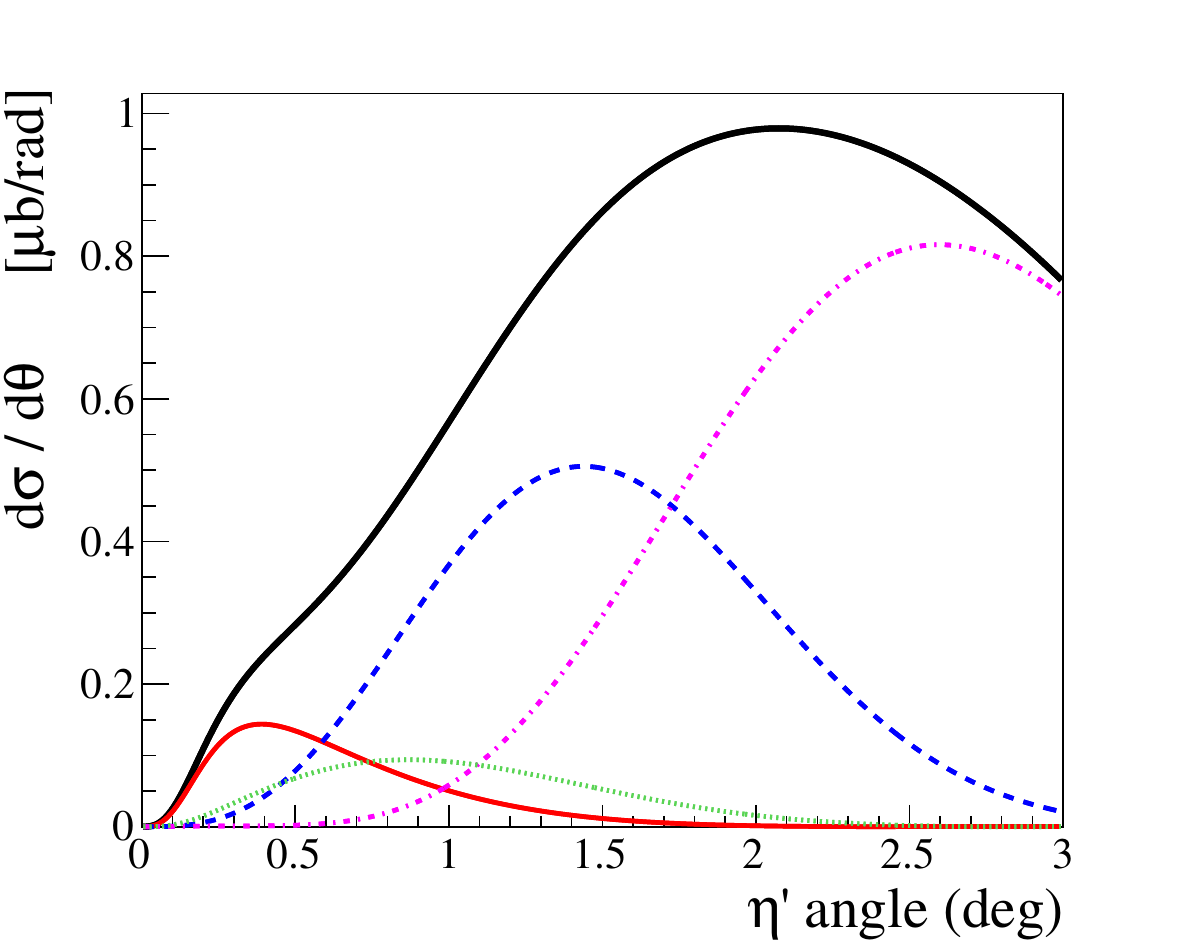} 
\includegraphics[width=0.48\linewidth,angle=0]{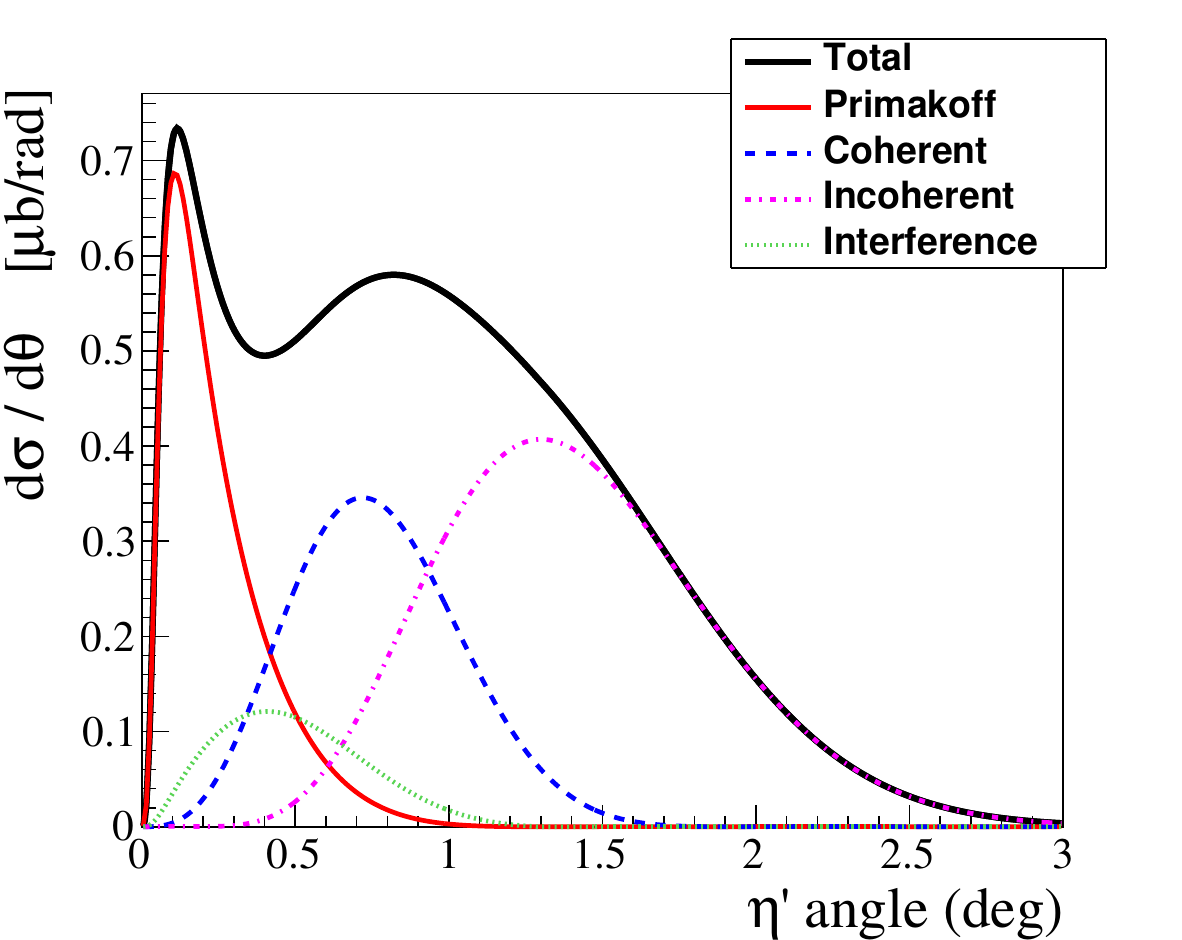}
\end{center}
\caption{The differential cross section for the production of $\eta^\prime$ mesons on a ${}^4$He target as a function of the production angle for beam energies of 10 GeV (left) and 20 GeV (right). The curves in this plot correspond to different physics processes, as explained in the text.}
\label{fig:etap}
\end{figure}

\begin{figure}[t]
\begin{center}
\includegraphics[width=0.75\linewidth,angle=0]{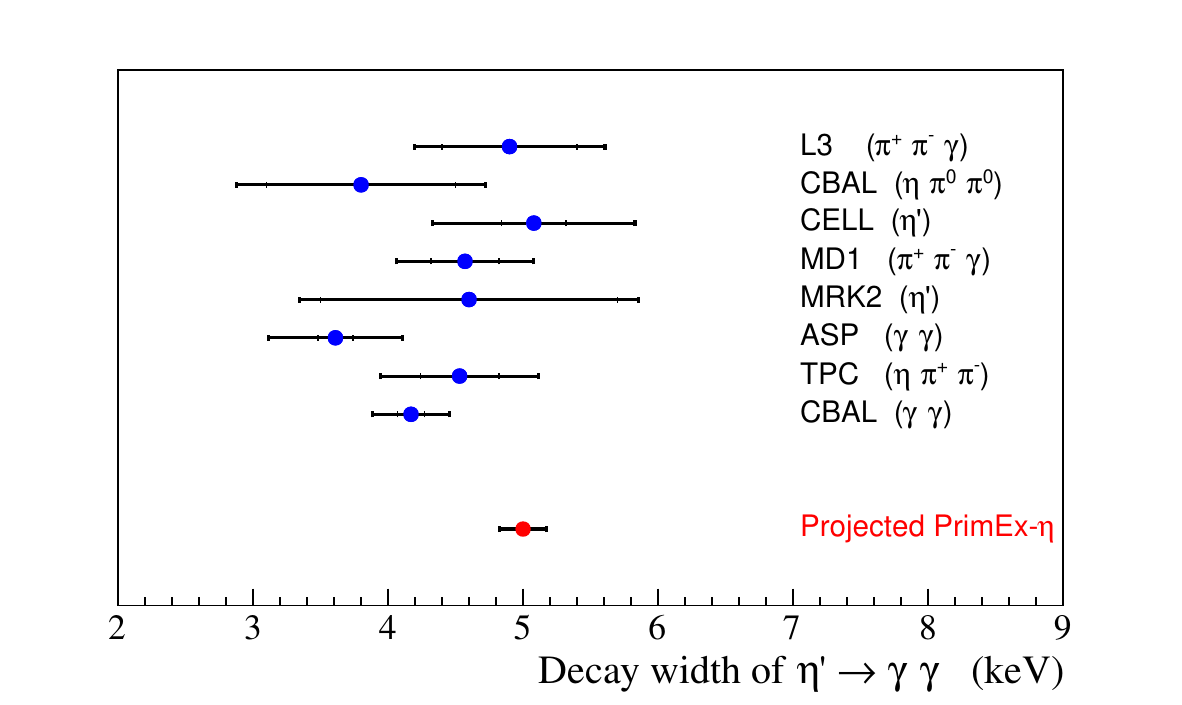} 
\end{center}
\caption{Blue circles: measurements of the $\eta^{\prime} \to \gamma \gamma$ decay width by collider experiments~\cite{ParticleDataGroup:2024cfk}. Red circle: projected error on the  $\eta^{\prime}$ decay width measured in the Primakoff process by the PrimEx $\eta$ experiment in Hall D using the possible 20 GeV CEBAF upgrade.} 
\label{fig:proj}
\end{figure}

The energy upgrade of the Jefferson Lab’s facility to 22 GeV~\cite{Accardi:2023chb} will allow for the extension of the Primakoff experimental program in Hall D. With an upgraded forward calorimeter and the future beam energy upgrade, we will significantly improve recent measurements of the PrimEx $\eta$ experiment of the $\eta$ differential cross sections at forward angles. This cross section will be used for the extraction of the $\eta$ decay width. The energy upgrade will allow to perform the first hight-precision measurement of the $\eta^\prime$ differential photoproduction cross section and to extract the radiative decay width via the Primakoff effect, which will complement the existing results from the collider experiments. Primakoff production of $\eta^{(\prime)}$ meson at larger energies has certain advantages: 
\begin{figure}[t]
\begin{center}
\includegraphics[width=0.75\linewidth,angle=0]{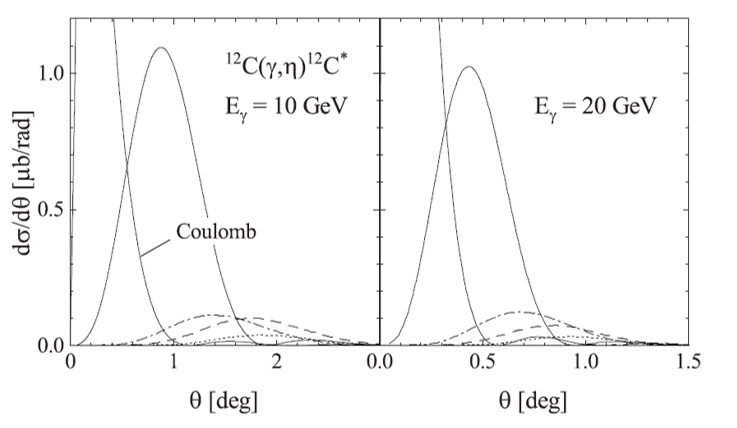}
\end{center}
\caption{
The differential cross section for the ${}^{12}C(\gamma,\eta)^{12}C^\star$ reaction computed for two beam energies: 10 GeV (left) and 20 GeV (right). The leftmost curves in these plots correspond to the Primakoff contribution (labeled "Coulomb"), the solid curves in the middle show the coherent background, and various excitations of the 1$p$-shell levels are depicted as dashed curves. The plot is taken from Ref.~\cite{Fix:2023vzj}.}
\label{fig:alex_fix}
\end{figure}
\begin{itemize}
\item the Primakoff cross section increases with the increase of the beam energy, 
\item the Primakoff and hadronic processes can be better separated at larger energies. The average angles of $\eta^{(\prime)}$ produced in  the Primakoff process and nuclear coherent (NC) background depend on the beam energy as $ \theta_{\rm {Primakoff}} \sim {m_{\eta^{(\prime)}}^2}/{(2\cdot E^2)}$ and $\theta_{NC}  \sim 2 / {(E\cdot A^{1/3})}$, respectively. As an example, the differential cross sections of the production of $\eta^{\prime}$ mesons on a ${}^4$He target for the 10 GeV and 20 GeV beam energies are shown in Fig.~\ref{fig:etap}. The cross sections were computed in the theoretical framework described in~\cite{Gevorkyan:2009mh},
\item the reconstruction of $\eta^{(\prime)}$ mesons in the Primakoff reaction   is significantly improved. According to the Geant detector simulation performed for the PrimEx $\eta$ experiment at 9 GeV, as well as the future experiment with the upgraded calorimeter and a 20 GeV beam energy, the polar angle resolution for $\eta \to \gamma \gamma$ decays is expected to improve from approximately $0.08^\circ$ to $0.02^\circ$. The simulation also predicts significant improvements in the resolution of the two-photon invariant mass, from 23 MeV to 14 MeV, and in the energy resolution of the reconstructed events, from about 300 MeV to 180 MeV. Similar improvement is also observed for $\eta^{\prime}$ decays, especially for $\eta^\prime \to \gamma \gamma$  and  $\eta^\prime \to \pi \pi \eta$ ($\eta \to \gamma \gamma$) decay modes,
\item in the Primakoff production the momentum transfer to the nucleus is getting  smaller at large energies. This may reduce uncertainties from nuclear excitations. The feasibility of using heavy targets is currently under study.
\end{itemize}
In the future experiments, $\eta^\prime$ mesons can be reconstructed using several decay modes.We estimated that the  $\eta^{\prime}\to\gamma\gamma$ decay width can be measured with an accuracy of about 3.5 $\%$~\cite{Accardi:2023chb} for one month of data taking on a $^{12}C$ target. The decay width measured by collider experiments and the projected errors of the future Primakoff experiment are shown in Fig.~\ref{fig:proj}.
 
\section{Summary}
The PrimEx $\eta$ experiment successfully completed data collection in 2022. The first physics publication, focusing on the measurement of the Compton scattering cross section in the energy range between 6.5 GeV and 11 GeV on a Be target, is currently in preparation. The analysis of the $\eta$ photoproduction cross section is underway, pending completion of the calibration and processing of the acquired data. This cross section will be used to extract the decay widths of $\eta\to\gamma\gamma$ and $\eta\to3\pi$. The upgrade of the GlueX forward calorimeter with high-granularity, high-resolution PbWO$_4$ crystals, along with recent theoretical calculations of nuclear excitations for heavy targets, stimulates new Primakoff experiments with the GlueX detector. The possible energy upgrade of Jefferson Lab's facility to 22 GeV would be beneficial for the extension of the Primakoff experimental program in Hall D.
\section{Acknowledgments}
This work was supported by the Department of Energy, USA. Jefferson Science Associates, LLC operated Thomas Jefferson National Accelerator Facility for the United States Department of Energy under contract DE-AC05-06OR23177.

\end{document}